# Emerging 2D Materials for Beyond von Neumann Computing: A Perspective

Yaser Banad

*School of Electrical and Computer Engineering, University of Oklahoma, Norman, OK 73019, USA*

**Abstract.** *The end of conventional Dennard scaling and the widening gap between memory bandwidth and arithmetic throughput have made the von Neumann partition a structural bottleneck rather than a transient one. Two-dimensional (2D) materials, with atomically thin geometries, electrically tunable carrier densities, and large optical responses, offer a unified platform on which to build devices that compute where they store, process events rather than clock cycles, and shift workload into the optical domain. This perspective surveys progress along three converging thrusts, graphene and graphene nanoribbon transistors as scalable channel materials, oxide and 2D-integrated memristors for in-memory analog compute, and silicon-compatible 2D photonic and thermal-emitter structures for optical computing primitives. Our central argument is that the 2D-materials community has spent a decade producing record devices, and the next decade will be decided by who first integrates three of them on a single semiconductor wafer.*

## 1. Introduction

Decades of CMOS scaling delivered exponential improvements in compute density, but the underlying von Neumann separation of memory and logic has not scaled with it. Modern accelerators spend the majority of their energy moving data, not transforming it, and the trend is worsening as model sizes outpace on-chip memory [1]. Three architectural responses have emerged in recent years: in-memory and resistive computing, neuromorphic and event-driven processing, and integrated photonics [2,3,4,5,6,7]. Each of these directions demands devices that bulk silicon does not natively provide: tunable, high-mobility channels at sub-10 nm pitches; non-volatile, analog-programmable conductances; and on-chip optical elements with reconfigurable spectral response.

Two-dimensional materials are uniquely positioned across all three needs. Graphene supports high carrier mobility and strong electrostatic gate control even at sub-5 nm widths, where bulk silicon channels fail [8,9,10]. Engineered transition metal oxides, especially when stacked with 2D contact and barrier layers, can be tailored into reliable analog memristors with controlled switching dynamics and predictable retention [11]. Multilayer graphene stacks act as electrically tunable mid-infrared emitters and absorbers, providing photonic primitives that are co-fabricable with electronic devices [12,13,14]. The remainder of this perspective examines each of these threads, then turns to the comparative landscape, the co-design and integration questions, and the concrete grand challenges that now dominate progress.

Why this perspective and why now. Three enablers that did not exist five years ago have converged. Wafer-scale CVD of monolayer and bilayer graphene with controlled domain orientation has matured to 200 mm and beyond [15]. Atomic-layer deposition of high-k dielectrics on graphene now routinely reaches sub-1 nm equivalent oxide thickness without unrecoverable mobility loss. And the first integrations of 2D-channel transistors into the back-end-of-line of advanced CMOS are being demonstrated by industry consortia and academic foundries, supported by recent advances in uniform wafer-scale epitaxy of bilayer transition metal dichalcogenides [16]. Each enabler removes a specific blocker that had previously confined 2D devices to small-area, transferred demonstrations. Together they lower the activation energy for the integration step that this perspective argues is the rate-limiting one. Figure 1 places the three architectural responses on a single map and shows where 2D materials act as a shared device substrate across all three.

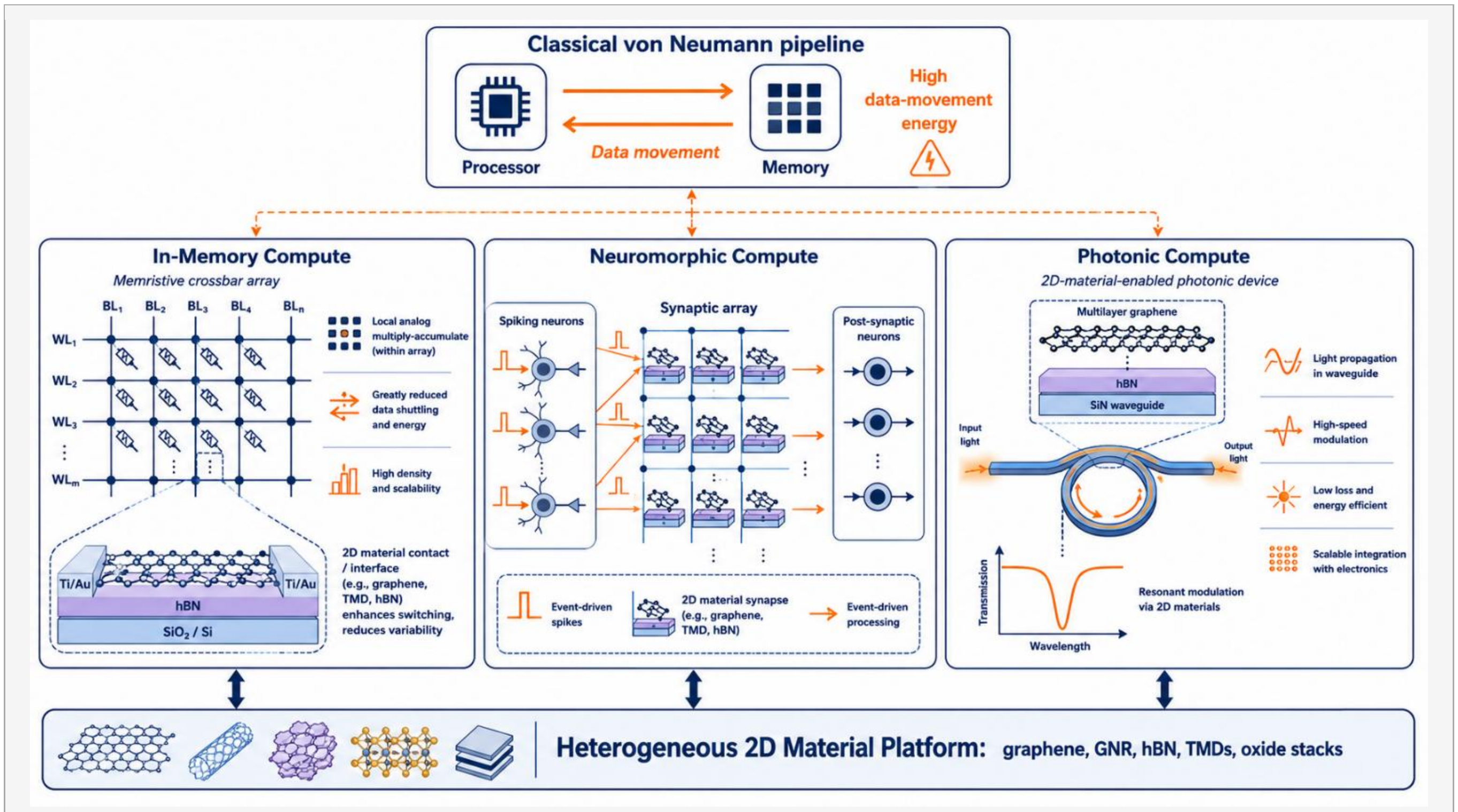

***Figure 1.*** *Architectural map of beyond von Neumann computing paradigms (in-memory, neuromorphic, photonic) and the role of 2D materials as a shared device substrate across all three.*

## 2. 2D Channels: Graphene and Graphene Nanoribbon Transistors

Pristine graphene is a semimetal and cannot directly support digital logic, which requires high on-off current ratios. Patterning graphene into nanoribbons (GNRs) opens a width-dependent bandgap and restores transistor-like switching behavior. Atomistic non-equilibrium Green function (NEGF) modeling has mapped this design space in detail: ribbon widths in the 1 to 3 nm range produce bandgaps comparable to InGaAs while preserving mobility advantages over silicon [9,10,17]. These models also clarify how source-to-drain tunneling, contact resistance, and ballistic transport regimes interact at scaled gate lengths, providing the framework needed for circuit-level projections.

Two practical limits dominate measured device performance. First, line-edge roughness on the nanoribbon couples to backscattering, degrading the on-current and broadening the subthreshold swing. Statistical NEGF simulations show that an rms edge roughness on the order of 0.5 nm can reduce on-current by tens of percent and shift threshold-voltage distributions enough to compromise large arrays [18]. Second, dielectric integration determines both transport, through remote phonon and Coulomb scattering from interface traps, and reliability. High-k dielectrics such as HfO2 and Al2O3 deposited by atomic layer deposition with appropriate seed strategies preserve channel transport while supporting equivalent oxide thickness below 1 nm, and the choice of dielectric is now understood to be inseparable from the choice of channel [19]. Figure 2 illustrates the GNR FET architecture together with the two limits described above: the rough nanoribbon edge at the channel level and the high-k gate stack, with an inset showing how $I_D$-$V_{GS}$ characteristics degrade as edge roughness increases.

Beyond logic, the strong gate dependence of GNR FETs has been exploited as a sensor primitive: scaled GNR FETs serve as nanometer-footprint on-chip thermometers, enabling dense local temperature monitoring in modern integrated circuits without a dedicated sensor process [20]. Compendium treatments of these design considerations, spanning materials, devices, and circuits, have been published [9].

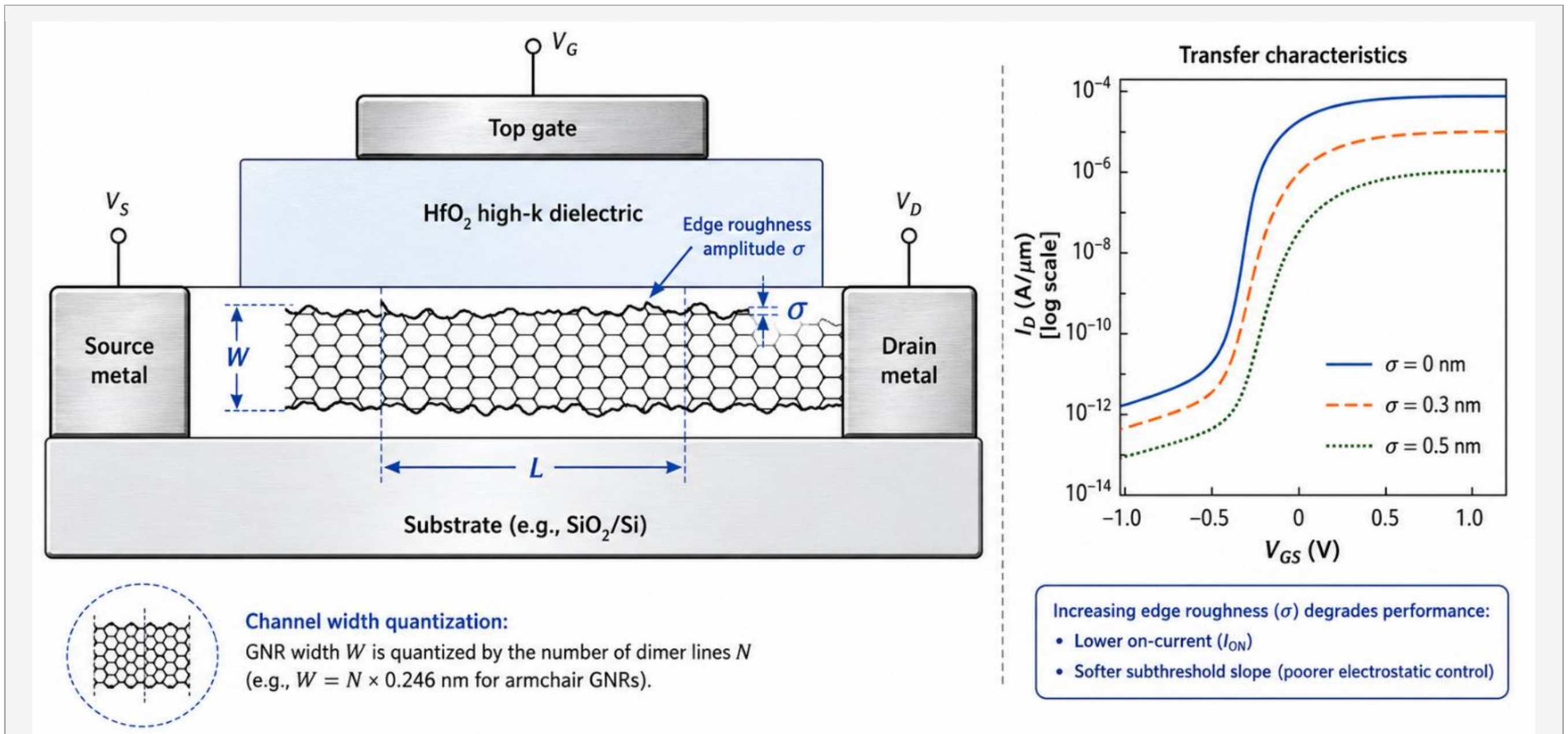

***Figure 2.*** *Graphene nanoribbon FET architecture showing channel width quantization, line-edge roughness, and high-k gate stack; inset shows simulated $I_D$-$V_{GS}$ sensitivity to edge roughness amplitude.*

## 3. Resistive Switching and Memristive 2D Stacks

Memristive devices implement the analog conductance updates required for in-memory matrix-vector multiplication, the elementary operation underlying both classical signal processing and modern neural networks [2,3]. The open question for hardware ML is no longer whether oxide stacks can switch resistively, that is well established, but whether their conductance trajectories, retention, and cycle-to-cycle variability are sufficient to encode deep-learning weights at useful precision.

Recent multiphysics analyses of TaOx and HfOx stacks reveal that the interplay between Joule heating in the conductive filament, oxygen vacancy migration, and electrode-oxide interfaces sets a fundamental tradeoff between switching speed, energy, and endurance [11]. Coupling thermal and electrical transport in a single simulation loop, rather than treating them sequentially, exposes degradation modes that isothermal models miss, including filament thinning under repeated programming and asymmetric SET and RESET dynamics. These insights translate directly into device-level design rules: filament confinement, electrode work function selection, and thermal boundary engineering.

Two-dimensional materials enter this picture in two complementary roles. As ultrathin tunneling barriers, hexagonal boron nitride and selected TMD monolayers can suppress current creep at low conductance states and improve the linearity of analog programming. As electrodes, graphene confines the active filament region and reduces parasitic capacitance, enabling sub-nanosecond pulse programming at modest write energies. The combination of validated electro-thermal models [11] and 2D-stabilized stacks is a credible path to memristive arrays with more than 100 distinguishable conductance levels suitable for inference-grade neural-network accelerators. Figure 3 shows a representative oxide memristor with 2D-material contact and barrier layers alongside the pinched I-V hysteresis loop and the electro-thermal coupling between Joule heating in the filament and oxygen-vacancy migration that together govern device behavior.

A practical recommendation for the field follows from the variability picture. Reporting median switching figures of merit, while convenient, hides the tails of the distributions where deep-learning accuracy actually breaks. Future memristor characterization work should report full conductance distributions, retention statistics across the operating temperature range, and bit-error-rate trajectories under representative workload pulse patterns, treating variability as a first-class design parameter rather than a defect to be averaged out [11]. The same convention should propagate to neuron and synapse circuits constructed from 2D-compatible primitives [21,22].

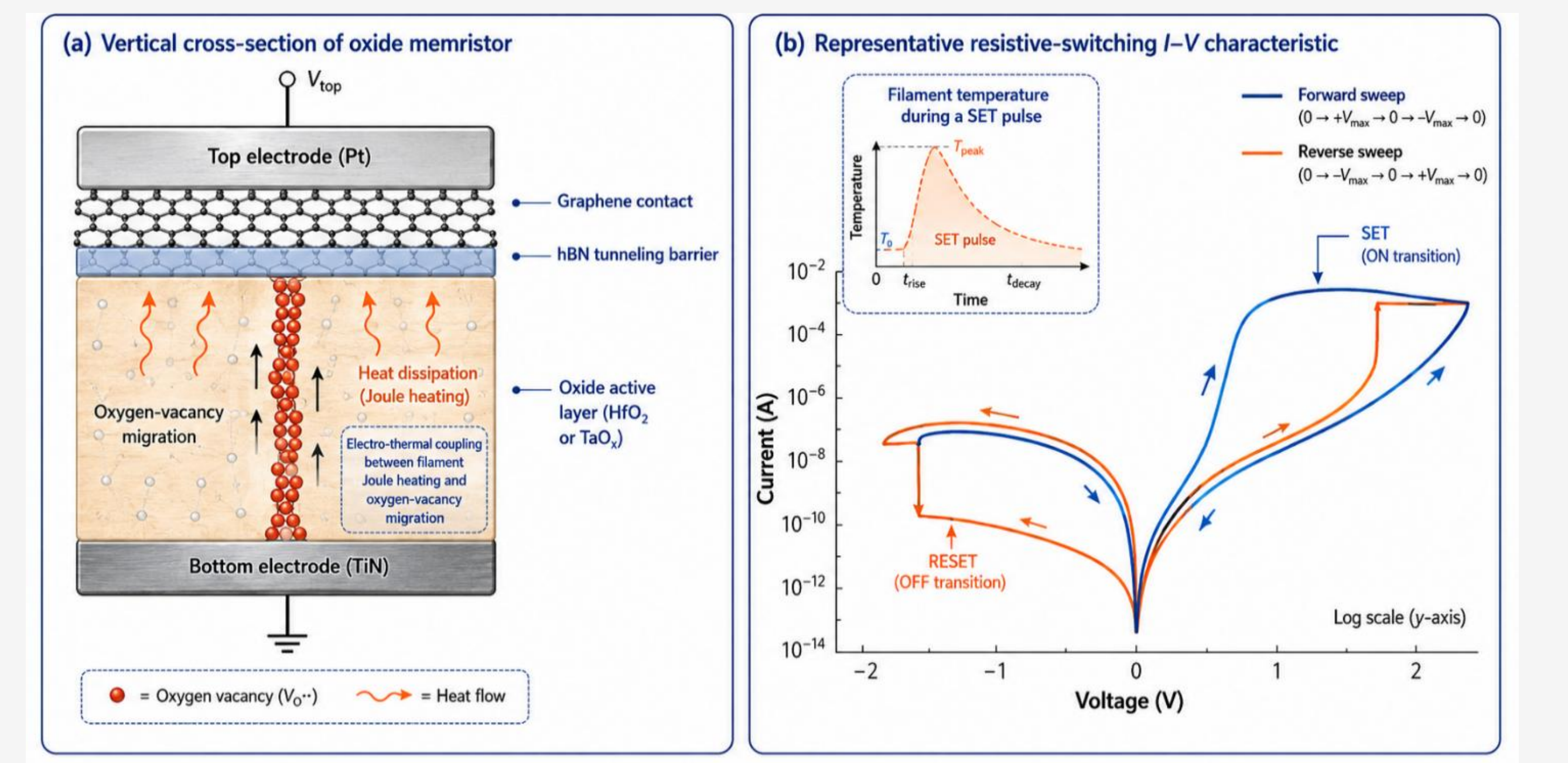

**Figure 3.** Oxide memristor with 2D-material contact and barrier layers; representative I-V hysteresis loop and electro-thermal coupling between filament Joule heating and oxygen-vacancy migration.

## 4. Neuromorphic Circuits and Event-Driven Compute

Spiking neural networks substitute discrete events for clocked floating-point operations and dramatically reduce dynamic energy when input activity is sparse [4]. Building efficient spiking hardware requires compact, low-energy neuron circuits that integrate inputs, generate spikes on threshold crossing, and reset, all while preserving compatibility with standard digital design flows.

Conventional CMOS integrate-and-fire (IF) neuron implementations in mature technology nodes, including 22 nm bulk and FDSOI, achieve picojoule-per-spike energies but at the cost of a sizable footprint [21]. Comparative analyses across topologies show that the energy, area, and maximum spiking-frequency Pareto front depends strongly on whether the neuron is constructed from purely transistor-based RC integrators or hybridizes diode-like and capacitive elements [21]. Replacing key components of the IF circuit with side-contacted field-effect diodes (S-FEDs) and other beyond-CMOS primitives has been shown to reduce both per-spike energy and silicon area while preserving compatibility with standard digital processes [22]. Figure 4 shows a compact S-FED-based integrate-and-fire neuron driving a 2D-material memristive synaptic crossbar, capturing the combination of beyond-CMOS neuron and 2D-stabilized synapse on which the integration argument of this perspective relies.

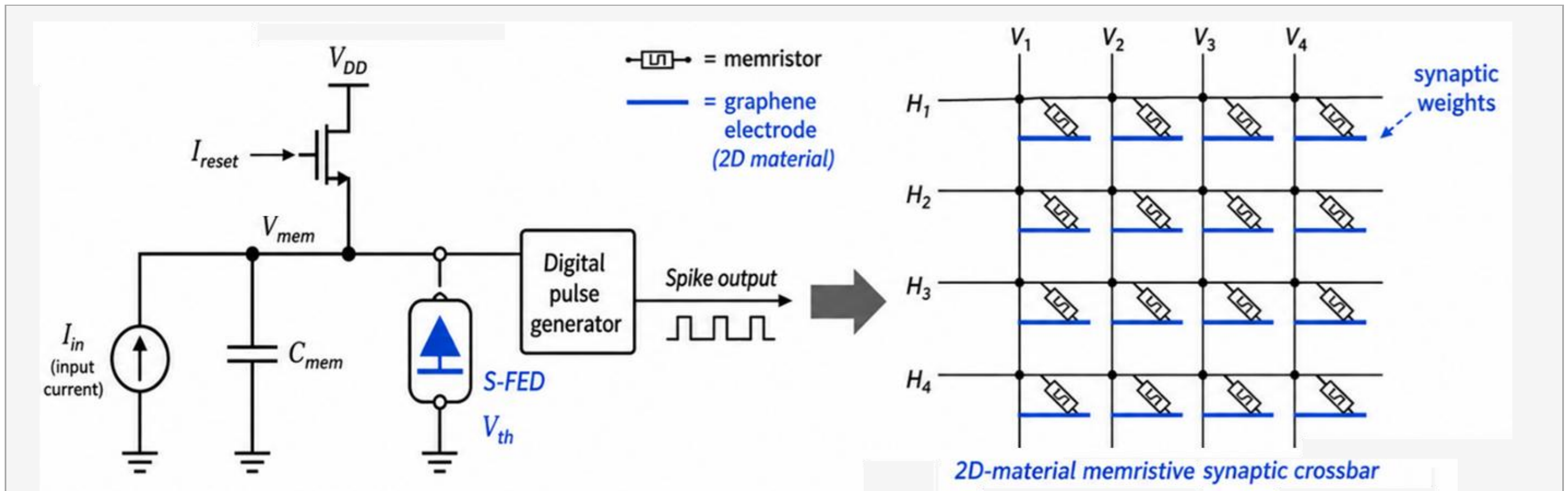

***Figure 4.*** *Compact integrate-and-fire neuron circuit using a side-contacted field-effect diode in place of a conventional comparator stage, driving a 2D-material memristive synaptic crossbar.*

The natural extension is to substitute the synapse with a 2D-compatible memristor [11], unifying weight storage, weight update, and neuron dynamics on a shared substrate. Such a unification is precisely what motivates a 2D-material-centric integration strategy: synapse, neuron, and interconnect, each sized correctly for its role, drawn from a shared device library.

## 5. 2D Photonic and Thermal-Emitter Primitives

A complementary route to beyond von Neumann compute encodes information in photons rather than electrons. Two opportunities are worth highlighting in the 2D-material context. First, multilayer graphene stacks placed on dielectric substrates support electrically and thermally tunable mid-infrared emission spectra. Aperiodic stacking sequences, including Cantor- and Fibonacci-like layer arrangements, generate engineered photonic states that produce switchable, narrowband thermal emission, with applications ranging from spectrally selective infrared sources to physical-layer security and on-chip thermal signaling [12,13]. The same physical mechanism, voltage-controlled modulation of graphene's optical conductivity, underwrites tunable photonic activation functions and weighted-sum elements in optical neural networks. Figure 5 illustrates this geometry, showing an aperiodic multilayer graphene stack on a dielectric substrate together with the gate-tunable narrowband emissivity spectra that make such structures useful as photonic compute primitives.

Second, 2D semiconductors and graphene-based modulators offer the bandwidth and electro-optic tuning range needed for matrix-vector multiplication at the speed of light, complementing larger-scale silicon photonic neural networks demonstrated in recent years [6,7]. Together, tunable graphene emitters and 2D-modulator-based photonic compute close the loop between in-memory analog compute and integrated photonic compute: both seek to amortize the cost of data movement, and both benefit from co-fabricable 2D building blocks.

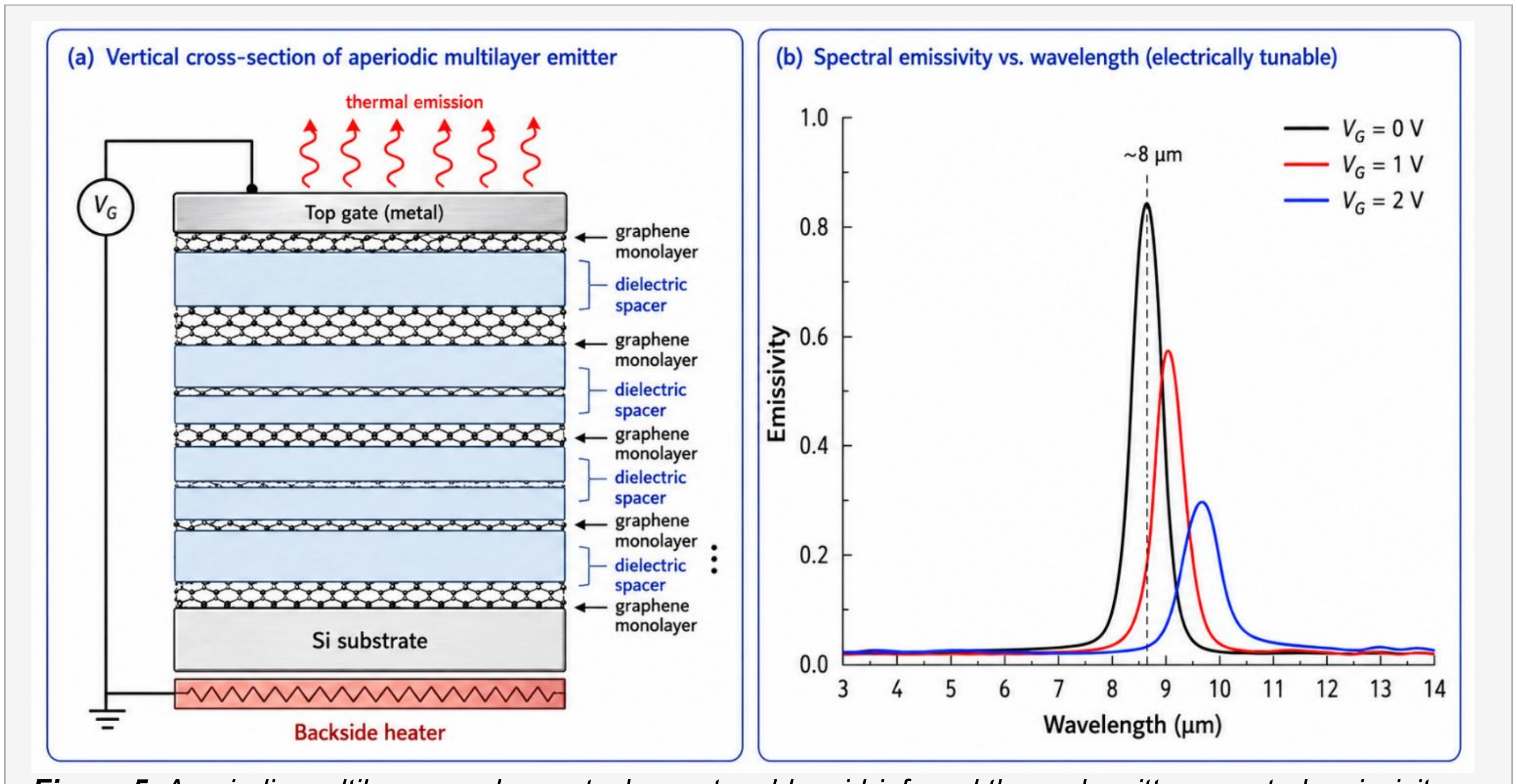

***Figure 5.*** *Aperiodic multilayer graphene stack as a tunable mid-infrared thermal emitter; spectral emissivity versus wavelength under different gate biases shows narrowband, electrically switchable emission.*

## 6. Comparative Landscape: 2D Materials Among Beyond-CMOS Options

Beyond-CMOS device research has produced several mature non-volatile memory and compute primitives, including phase-change memory (PCM) [23], ferroelectric devices and FeFETs [24], spin-transfer-torque magnetic memory (STT-MRAM) [25], and 2D-material-based memristors and transistors [11,18,19]. Each option occupies a distinct corner of the energy-density-endurance-precision tradeoff space. Table 1 summarizes the comparison along axes most relevant to in-memory and neuromorphic compute. The table is intended as honest accounting rather than ranking: 2D materials are not strictly best on any single axis.

What this comparison makes visible is that no single non-2D platform offers all four of analog precision, density, endurance, and a native photonic mode at once. PCM and FeFET dominate analog memory on endurance and density. STT-MRAM dominates endurance but is essentially binary. Only 2D materials, taken as a family, span the full set of compute primitives, including the optical-electronic interface that all-electronic options simply cannot supply [12,13,14]. The cost of this breadth is that 2D platforms remain less mature on any individual axis. The strategic question for the field is therefore not whether a 2D memristor can outperform PCM in isolation, but whether a 2D heterogeneous chip can outperform a heterogeneous system that bolts PCM, FeFET, and silicon photonics together from separate process flows. The integration argument of this perspective is exactly that the 2D path wins on integration cost, even if it loses individual cells.

| Technology | Energy/op | Density | Endurance | Analog levels | BEOL compatibility | Native photonic mode |
|---|---|---|---|---|---|---|
| **2D-material memristors / FETs** | Sub-pJ | High | 10^6-10^9 | 4-6 bit | Strong | Native (graphene, TMD) |
| **Phase-change memory (PCM)** | pJ | High | 10^9-10^10 | 4-6 bit | Strong | Limited (waveguide-PCM hybrids) |
| **Ferroelectric / FeFET** | fJ-pJ | Moderate | 10^10-10^12 | 2-4 bit | Strong | None |

| Technology | Energy/op | Density | Endurance | Analog levels | BEOL compatibility | Native photonic mode |
|---|---|---|---|---|---|---|
| **Spintronic / STT-MRAM** | Low | Moderate | >10^15 | 1-2 bit | Strong | None |
| **Si CMOS (digital reference)** | Low | High | Volatile | Digital (high) | Native | Modulators only |

***Table 1.*** *Comparative landscape of beyond-CMOS compute primitives. Entries are typical orders of magnitude rather than records. The 2D-material row is highlighted to indicate the family argued for in this perspective; the comparison is intended as honest accounting, not as a ranking.*

## 7. Algorithm-Hardware Co-design

Hardware progress alone will not deliver the projected energy and accuracy gains; algorithm and software stacks have to budget explicitly for analog non-idealities. Off-the-shelf neural network training, when deployed naively on memristive crossbars, loses several points of accuracy from a combination of conductance drift, asymmetric updates, and read noise [26]. Co-designed training procedures, including non-ideality-aware quantization, noise injection during training, and online retraining against measured device statistics, recover most of this gap and in some cases match digital baselines on equivalent power budgets [27].

Two implications follow for the 2D-material program. First, device-level reporting must include the parameters that algorithm designers actually need: per-cycle conductance trajectories, asymmetric SET and RESET characteristics, and temperature-dependent retention curves [11]. Second, the natural unit of evaluation is no longer a single device but the device-plus-mapping-plus-training-recipe stack. The next round of competitive benchmarks should compare 2D-material-based stacks against PCM and FeFET stacks at the level of complete inference workloads, not isolated device figures of merit. This shift turns hardware-software co-design from a soft recommendation into a measurable engineering discipline.

## 8. A Concrete Near-Term 2D-Material Demonstrator

To make the integration argument concrete, consider a near-term demonstrator chip that exercises all three pillars on a single substrate. The digital control plane is implemented in scaled GNR FETs [18,19], providing high mobility and electrostatic control at sub-10 nm pitch alongside on-chip thermometry from gate-tuned GNR sensors [20]. The compute core is a memristive crossbar with 2D-stabilized stacks [11], operated as an in-memory matrix-vector multiplier for the dense layers of a vision or language model. A spiking front-end built from S-FED neurons [21,22] handles event-driven preprocessing of camera or microphone input, gated into the analog crossbar only when activity is present. A photonic readout, based on aperiodic multilayer graphene emitters [12,13] integrated on the same wafer, provides electrically tunable mid-infrared output for inter-chip optical interconnect, sidestepping pin-bandwidth limits. Figure 6 sketches the floorplan of such a chip, with each pillar mapped to a specific block on a single CMOS-compatible substrate.

Each block has been demonstrated separately. None of them has been integrated together. The first team to do so on a single wafer with acceptable yield and a credible thermal budget [15,16] will define the reference platform for 2D-material-enabled beyond-von-Neumann compute, and the resulting chip will become the test vehicle on which co-designed training algorithms and architectural studies are evaluated for the rest of the decade.

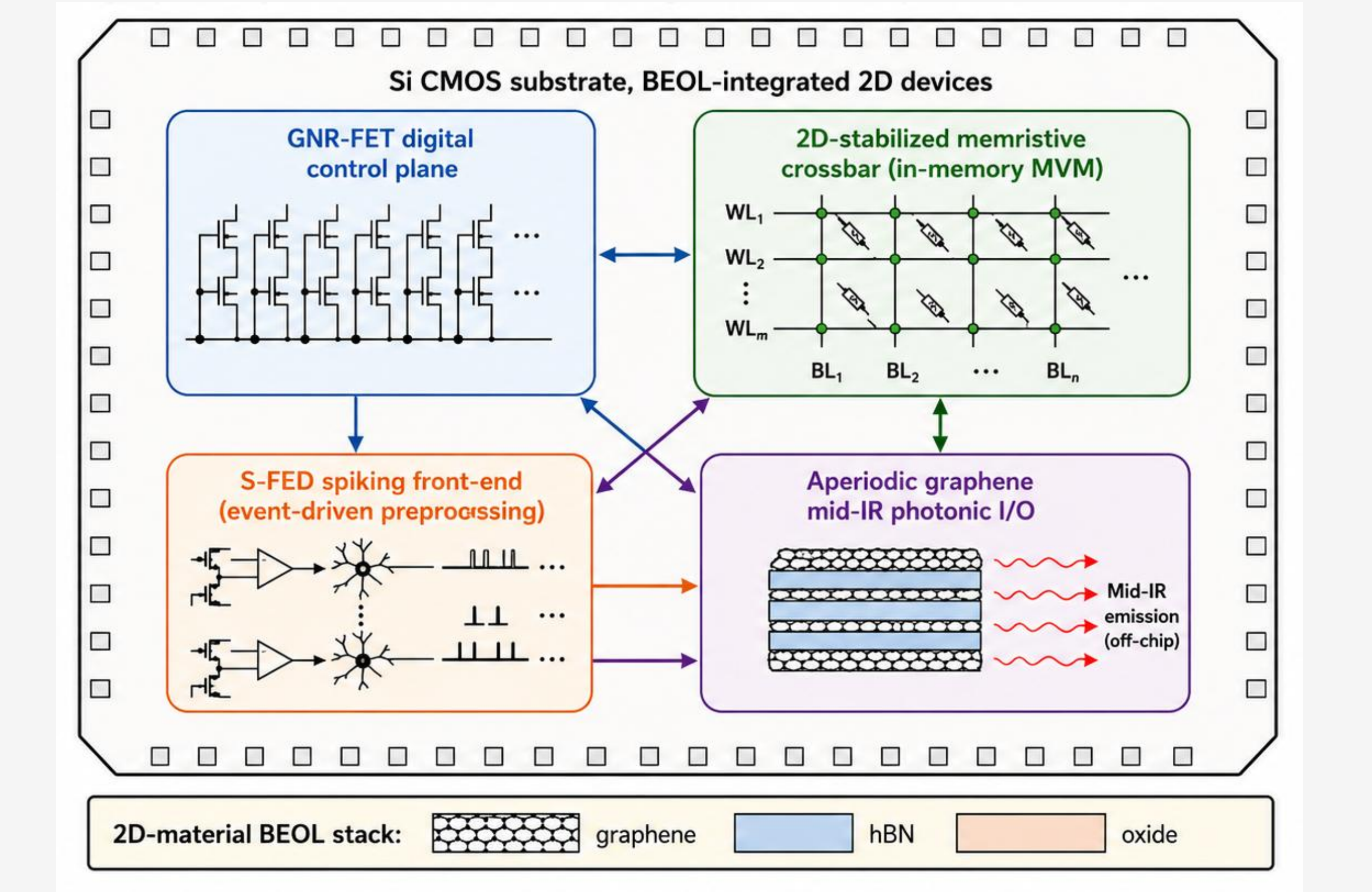

***Figure 6.*** *Conceptual layout of a near-term 2D-material heterogeneous compute chip integrating GNR-FET digital control, a 2D-stabilized memristive crossbar, an S-FED spiking front-end, and an aperiodic graphene photonic readout on a single CMOS-compatible substrate.*

## 9. Grand Challenges and Roadmap

The integration program implied by the previous sections decomposes into a small number of well-posed challenges, each measurable, each currently unresolved. Stating them in quantitative form is itself a contribution: it converts "more research is needed" into discrete milestones that can be funded, benchmarked, and contested. Figure 7 summarizes these targets in two panels: the left panel shows the layered integration stack on which the five challenges have to converge, each tagged with its target metric, and the right panel quantifies the gap between the current state of the art and the 2030 target for each challenge. The gap is exactly the integration program proposed in this perspective.

Figure 8 sketches a five-year roadmap that places these challenges on a single timeline across the three pillars, ending in the demonstrator chip of Section 8. The roadmap is deliberately aggressive: each milestone has a published precursor, but none has been delivered at the wafer scale and yield required for industrial pickup. The institutional teams that achieve this convergence will be those that combine material growth, device modeling, circuit design, and photonic integration as a single program rather than as adjacent disciplines. The technical pieces are in hand. What remains is the harder organizational and engineering task of putting them together at scale, and the field's strongest work in the next five years will come from groups that take that integration burden seriously.

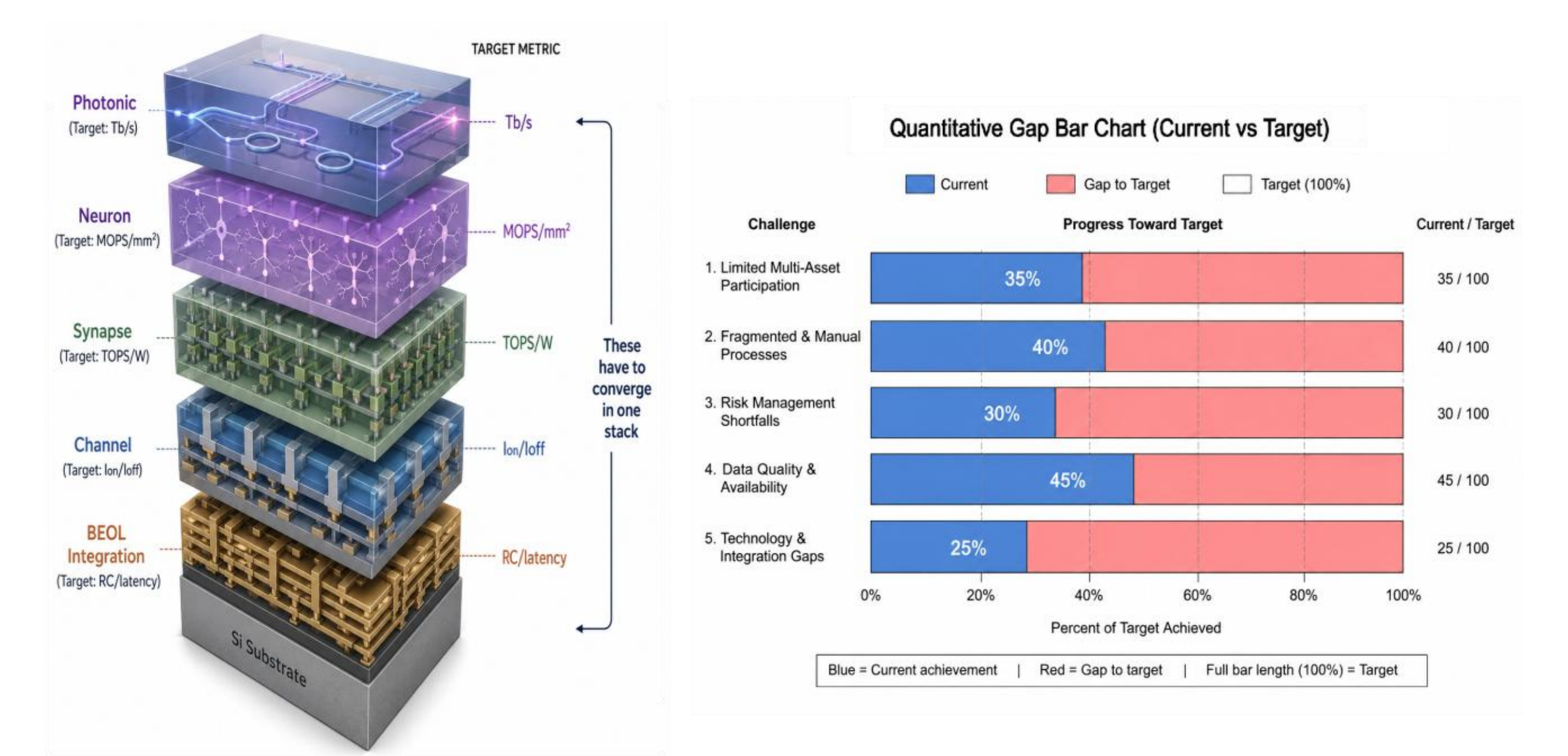

***Figure 7.*** *Two-panel summary of the Grand Challenges. Left: integration stack showing the BEOL, Channel, Synapse, Neuron, and Photonic layers that must converge on a single substrate, each annotated with its target performance metric. Right: quantitative gap chart comparing the current 2D-material state of the art (blue) with the 2030 target (red gap to target) for each of the five Grand Challenges. The visible gaps are the integration program proposed in this perspective.*

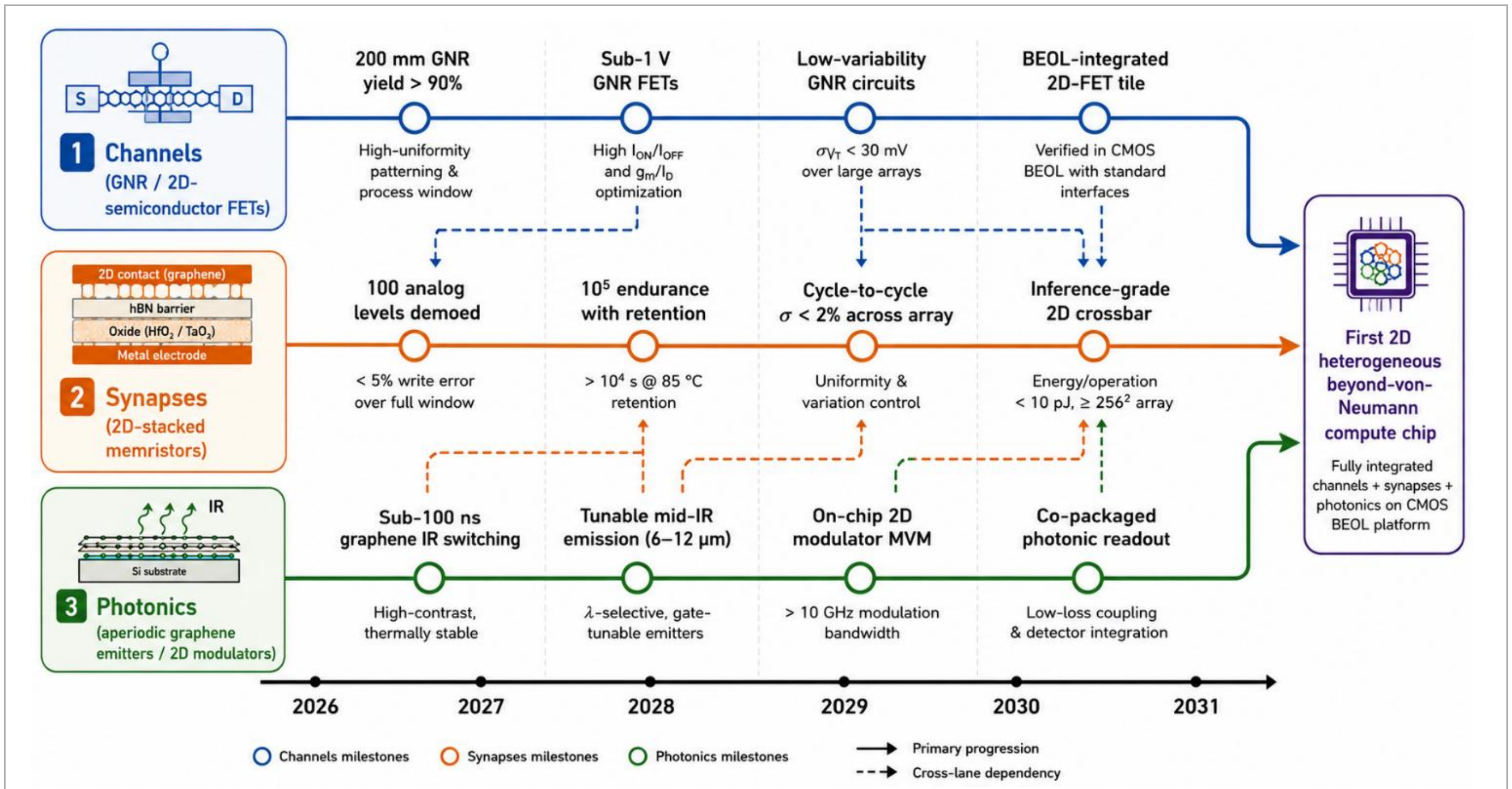

***Figure 8.*** *Five-year integration roadmap (2026 to 2031) for 2D-material beyond-von-Neumann compute, with milestones on three swim lanes (channels, synapses, photonics) converging on a heterogeneously integrated demonstrator chip.*